\documentclass[trackchanges]{aastex7}

\usepackage{makecell}
\usepackage{booktabs}
\usepackage{gensymb}
\usepackage{natbib}

\begin{document}

\title{The Near-Infrared Spectral Characteristics of Water Ice, Epsomite, and Halite Mixtures Relevant to Europa}

\author[orcid=0000-0000-0000-0001,sname='Berdis']{Jodi R. Berdis}
\affiliation{Johns Hopkins University Applied Physics Laboratory}
\email[show]{jodi.berdis@jhuapl.edu}  

\author[orcid=0000-0000-0000-0002,sname='Wagoner']{Carlie Wagoner} 
\affiliation{Johns Hopkins University Applied Physics Laboratory}
\email{carlie.wagoner@jhuapl.edu}

\author[orcid=0000-0000-0000-0002,sname='Takeuchi']{Akemi Takeuchi}
\affiliation{Johns Hopkins University Applied Physics Laboratory}
\affiliation{University of Maryland}
\email{atak@terpmail.umd.edu}

\author[orcid=0000-0000-0000-0002,sname='Hibbitts']{Karl Hibbitts}
\affiliation{Johns Hopkins University Applied Physics Laboratory}
\email{karl.hibbitts@jhuapl.edu}


\begin{abstract}

We publish the first NIR spectra of grain particulate mixtures of water ice, epsomite, and halite at cryogenic temperatures. Furthermore, we perform a quantitative assessment of the ability of both intimately- and linearly-mixed models to reproduce laboratory data of different grain mixtures of water ice, as well as water ice mixed with epsomite. We find that smaller grains of water ice impart a stronger influence than larger grains of water ice on the $2.0$ $\mu$m spectral feature in epsomite, and grain size signatures for both halite and epsomite are challenging to discern for larger grain sizes as a result of the saturated absorption features. These findings may indicate that an observation bias toward smaller grain sizes of ice could exist, and that quantitative assessments provided by spectral mixture analyses will be the most reliable method for determining compositions and abundances of materials. We also find that the linearly-mixed and intimately-mixed models of water ice appear to match the laboratory spectra as expected, though still display some inconsistencies, often either in the continuum or the absorption features. When modeling pure water ice and water ice mixed with epsomite, no discernible difference is observed between the fits of the linearly- and intimately-mixed models. Future spectral mixture analyses that use epsomite should be aware of a potential error in the published epsomite optical constant data, in which the cryogenic data appear to be taken at ambient conditions.

\end{abstract}




\section{Introduction} 

Europa's surface is predominantly composed of water ice with varying amounts of hydrated material(s) that are believed to be indicative of the composition of Europa’s subsurface ocean \citep{pilcher+1972, consolmagno1975, mccord+1998}, and thus can provide insight into its potential habitability. The presence of water ice on or within an object is intriguing due to the possible habitability implications if that object possesses warm pockets containing liquid water, which is one of the requirements for the development and sustenance of Earth-like life. Many studies have conducted compositional analyses for Europa to investigate the spatial distribution of water ice and non-water ice species on the surface in order to identify processes that may be influencing their distribution, and whether these processes are endogenic or exogenic (e.g., \citealt{mccord+1998, carlson+1999}, etc). Understanding the properties of water ice on the surface, including its temperature and grain size distribution, is critical to better elucidate the effects of surface processes, as well as potential impacts of subsurface activity on the surface. Several non-water materials have been proposed to exist on the surface based on the presence of their optical and NIR spectral features, such as sodium and magnesium salts \citep{shirley+2010, brown+hand2013, ligier+2016, trumbo+2019, king+2022}. Epsomite (MgSO$_4$ $\cdot$ $7$H$_2$O) and halite (NaCl) are two possible hydration states of magnesium sulfate and sodium chloride that may be present on Europa’s surface, and we focus on these two materials, along with water ice, in this study.

This work is particularly relevant and timely in anticipation of the arrival of NASA's Europa Clipper mission and ESA's JUICE mission. These spacecraft will collect reflectance data from the visible (Vis) through the near-infrared (NIR) with Europa Clipper's Europa Imaging System (EIS; \citealt{turtle+2024}) and Mapping Imaging Spectrometer for Europa (MISE; \citealt{blaney+2024}), as well as JUICE's Moons and Jupiter Imaging Spectrometer (MAJIS; \citealt{poulet+2024}), between them yielding a wealth of spectral information about Europa’s surface. Accurate compositional characterization of water ice and non-ice species will require comparison with the spectral properties of these materials measured in the laboratory. Such spectral analog data are critical for maximizing the science return from the Europa Clipper mission through the use of spectral mixture analysis. 

The compositional interpretation of surface material using spectroscopy can be achieved through spectral mixture theory, the concept that a single observed spectrum is the combination of various pure material spectra, or endmembers, weighted by the respective fractional abundance that exists in that single observation, either as discrete, separate patches of materials or mixed at the grain scale. Spectral mixture analyses have been performed on hyperspectral image data for a majority of the solid planetary bodies in the solar system, including the Moon \citep{li+mustard2003}, Mars \citep{combe+2008, liu+2016}, asteroids \citep{combe+2015}, Saturnian satellites such as Titan \citep{lecorre+2009, mccord+2008}, and Jovian satellites such as Europa \citep{shirley+2010, dalton+2012}.

\subsection{Radiative Transfer Theory and Spectral Mixtures}

Optical constants, which describe the inherent properties of a material and its interactions with incident light, are critical to the accurate interpretation of reflectance spectra from planetary bodies when modeling granular, or `intimate', mixtures. Spectral mixture analysis uses reference library data, either spectra or optical properties, to infer the composition of a surface from remote sensing spectral data. Where linear mixtures combine reflectance spectra of each component, non-linear or intimate mixtures attempt to model how the components interact with each other. In this approach to intimate mixture modeling, we combine the single-scattering albedo of each material component, ensuring that the behavior of light when encountering different material grains is accounted for (e.g., \citealt{dalton+pitman2012}). Reference data can be laboratory spectra or model spectra generated using radiative transfer theory – an approach that typically requires optical constants for the reference material over the wavelength range of interest. Optical constants in concert with a bidirectional reflectance model allow for the simulation of all possible viewing geometries, as well as variations in the material’s physical characteristics and are thus the most effective technique for generating model spectra (e.g., \citealt{hapke2012}).

Hapke scattering theory provides a method for estimating the bidirectional reflectance of a material, including granular mixtures, by taking into consideration the scattering effects, viewing geometry, and real and imaginary indices of refraction for that material \citep{hapke1981}. When optical constant information has been produced for a material, it is possible to synthesize a reflectance spectrum of that material's presence on an airless solid surface given the regolith properties (e.g., porosity, grain size), temperature, illumination, and incident and emission angles of the synthesized observation. Hapke modeling has been frequently used in the literature to aid in performing spectral mixture analyses on airless solid surfaces in order to more accurately synthesize endmember spectra \citep{lucey1998, mustard+pieters1989, hamilton+2005, warell+davidsson2010, ciarniello+2011, dalton+2012}. When optical constant information does not exist for a given material, cryogenic laboratory reference spectra acquired at environmental conditions similar to those in which the observed materials exist may alternatively be used; however, properties of the cryogenic laboratory materials such as viewing geometry and grain size are less likely to match those of the observed materials. In this case, linear mixtures of reflectance spectra have been used to approximate the material abundances on the surface of Europa \citep{ligier+2016, king+2022,berdis+2022}, however, a quantitative assessment of the fidelity of linear mixtures compared with intimate mixtures of materials relevant to Europa’s surface has not yet been made.

\subsection{Previous Work on Water Ice, Epsomite, and Halite}

In the NIR, water ice absorption features are caused by the three fundamental vibration modes of the water molecule and the combinations and overtones of those fundamental modes (\citealt{wilson+1955, hobbs1974}, and references therein). Most notably, the $2.0$, $1.65$, and $1.5$ $\mu$m bands are caused by the complex combinations and overtones of the fundamental modes, and their shapes and positions have been well characterized as a function of grain size and temperature (e.g., \citealt{clark1981, grundy+schmitt1998, dalton+2005}). Magnesium sulfate does not possess any vibrational features in the NIR, but easily accepts water molecules; therefore the absorption features of hydrated forms of magnesium sulfate, such as epsomite, are mostly due to the absorption features brought on by the addition of water molecules, and therefore, the spectrum becomes more complex at higher hydration states (\citealt{deangelis+2017, dalton2003, dalton+2005}, and references therein). Halite is also spectrally featureless in the NIR, though the NIR spectrum of halite often displays a $1.95$ $\mu$m adsorbed water feature.

Finally, grain sizes of water ice, epsomite, and halite have not yet been constrained on the surface of Europa. The NIR spectra of water ice and epsomite are highly grain-size-dependent, or as for halite, act as a `gray body' and thus, a better understanding of the grain size distribution on the surface is necessary to ensure accurate retrievals of material abundances from spectral mixture analyses. Therefore, we include several grain size bins for each of these materials in this study to further explore how grain size influences the NIR spectra of these materials, and how abundance retrievals may be used to approximate which grain sizes are present.

Furthermore, while many previous studies (e.g., \citealt{vu+2016, cerubini+2022, fulvio+2023}) have prepared frozen salty aqueous solutions as a means of representing potential cryovolcanic emplacements, these briny solutions can make it challenging to discern the resulting hydration level of the salt. For example, epsomite consists of magnesium sulfate (MgSO$_4$) bound to seven water molecules; by mixing epsomite salt with liquid water, then freezing, grinding, and sieving the result, it is not clear what hydration level of magnesium sulfate would result, and comparison to modeled spectra would be fruitless. Additionally, recent studies have suggested that radiolysis has likely desiccated the halide salts that are present on the surface, for example, \citet{trumbo+2019} found anhydrous NaCl, rather than a water-NaCl mixture such as hydrohalite (NaCl $\cdot$ $2$H$_2$O) in Hubble Space Telescope data. We specifically choose to prepare our samples as mixtures of discrete grains of water ice, epsomite, and halite as this is likely more representative and relevant to Europa's surface.

\begin{figure*}[ht!]
\plotone{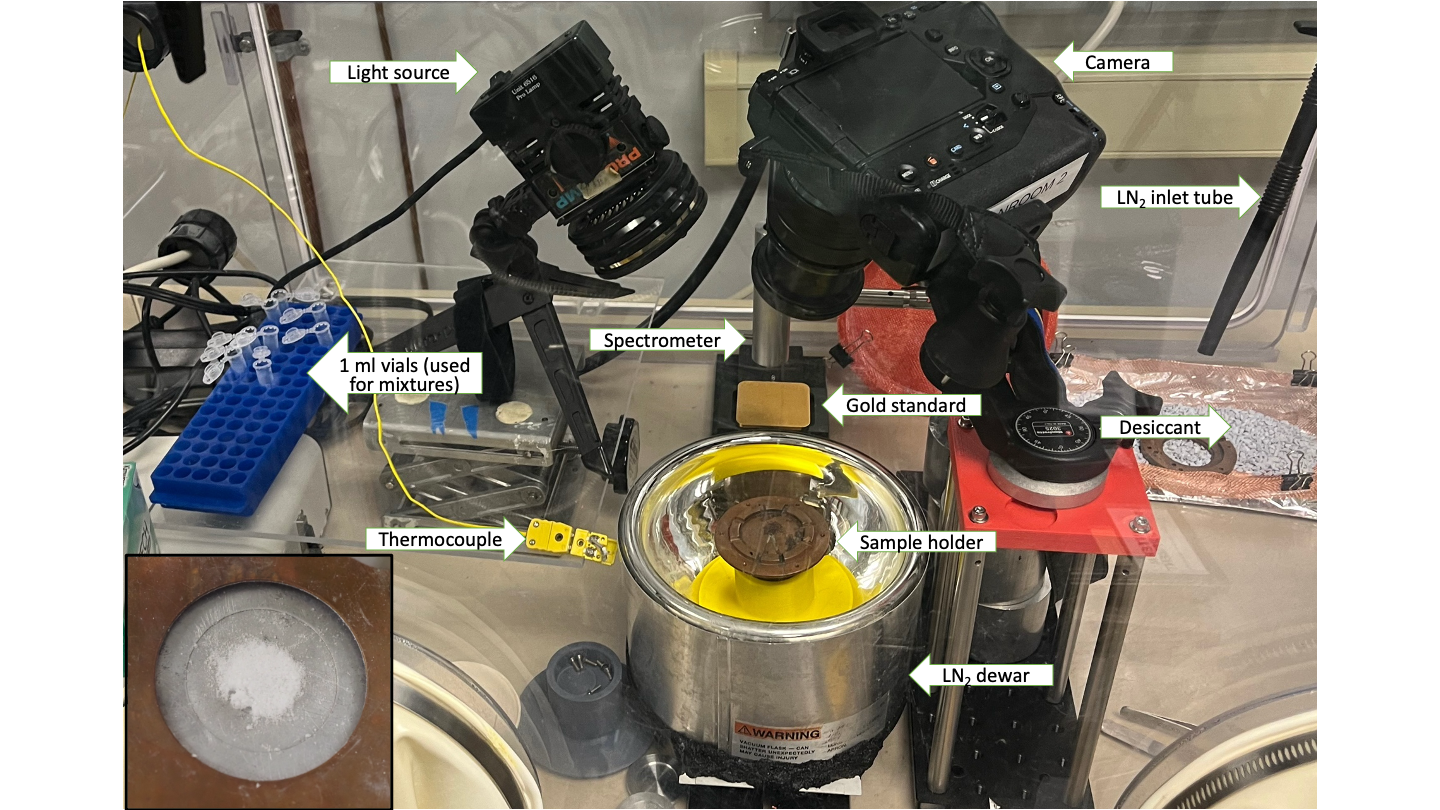}
\caption{Glove box setup for the experiments conducted in this study. The LN$_2$ dewar and gold standard are mounted on a linear stage that allow for the gold and sample to move into and out of the field-of-view of the spectrometer fiber-optic for measurements. Bottom left inset: photograph of 50\% $<63$ $\mu$m + $50\%$ $250-500$ $\mu$m water ice.
\label{fig:labsetup_diagram}}
\end{figure*}

\section{Methods}

\begin{figure*}[ht!]
\plotone{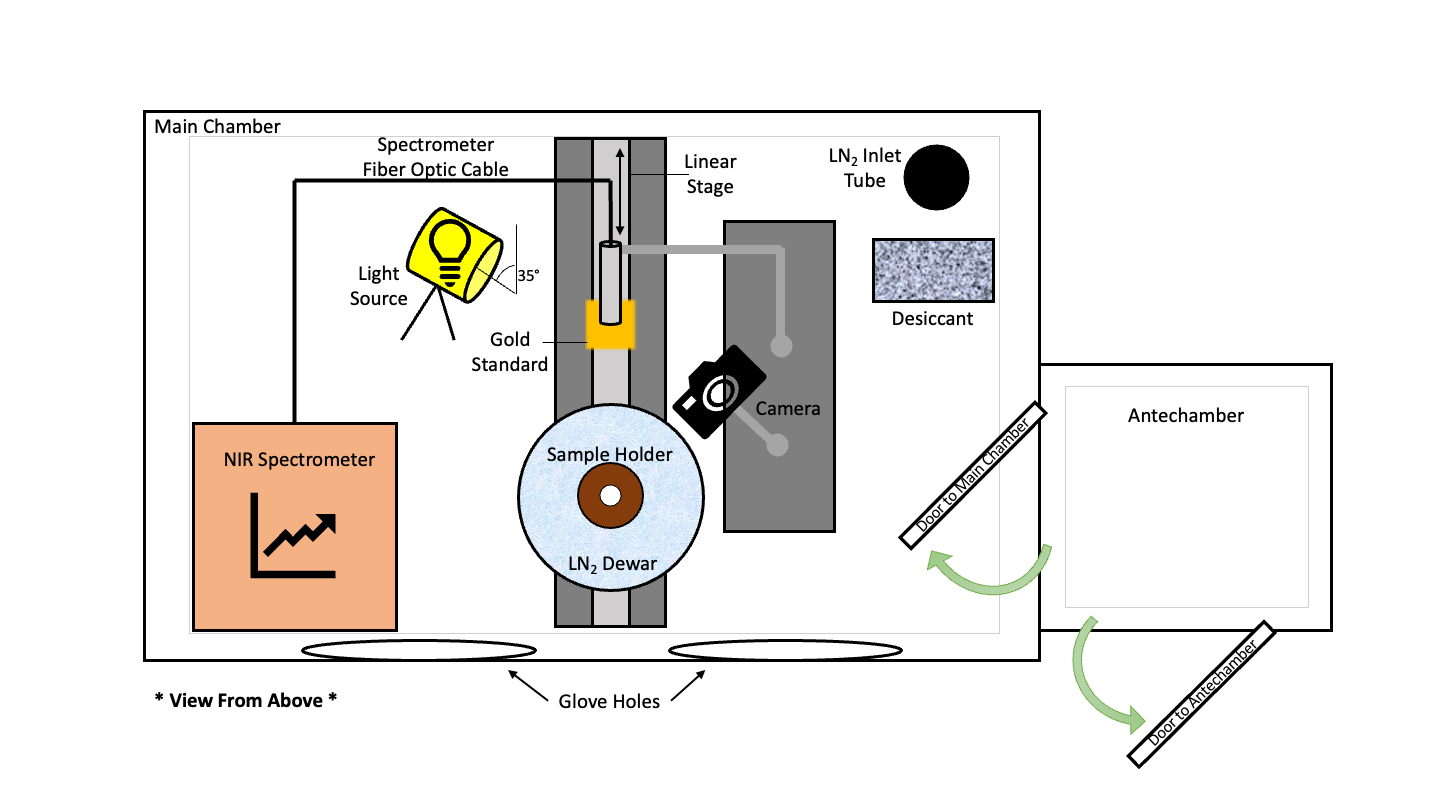}
\caption{Block diagram of the CRONOS laboratory setup; dimensions are not necessarily to scale. All materials and equipment were given at least 30 minutes in the antechamber before moving into the main chamber, to ensure no atmosphere was exchanged between the two areas.
\label{fig:glovebox_diagram_v1}}
\end{figure*}

We developed a state-of-the-art facility, the Cryogenic Reflectance Observations using Nitrogen-purged Optical Spectroscopy (CRONOS) facility, within an existing labspace at the Johns Hopkins Applied Physics Laboratory specifically for the experiments conducted in this study. Since we were handling water ice grains, an inherently cryogenic material, it was necessary to ensure that the water ice stayed frozen (and for relevance to Europa at cryogenic temperatures of $<120$ K), and also had as little exposure to water vapor as possible to avoid  the condensation of water frost from the atmosphere onto the grains that would otherwise contaminate the NIR spectra. This necessitated a nitrogen-purged glove box in which the samples were prepared, photographed, and measured spectroscopically. The glove box contained the incidence source at $35\degree$ from normal. The illumination was not collimated, so this represents an average incidence angle of illumination that may have spanned tens of degrees. While spectra were acquired quickly, it is likely that the sample still may have warmed slightly above the recorded temperature, up to $20$ K based on testing during which the thermocouple was placed within the sample pile. An increase of $20$ K would impact the shapes of the absorption features, notably the $1.65$ $\mu$m absorption feature in water ice, however we found that the alterations were not significant enough to impact the resulting model fits, as the changes were relatively small compared to the differentials between laboratory spectra and modeled fits - this is consistent with the temperature-dependent water ice spectra published by \citet{grundy+schmitt1998}. In fact, the depth of the $1.65$ $\mu$m band was indeed used here to confirm the temperature of the water ice, and extrapolated to the other materials. The fiber-optic cable attached to the spectrometer collected measurements at $0\degree$ from normal. The sample, sample holder, thermocouple, liquid nitrogen (LN$_2$) dewar, and a gold standard were also inside the purged glove box (Figures \ref{fig:labsetup_diagram} and \ref{fig:glovebox_diagram_v1}).

\subsection{Sample Preparation and Spectral Acquisition}

Whereas previous studies have often prepared samples by freezing salty aqueous solutions, we specifically chose to mix together discrete grains of materials, as this is able to be modeled and is likely more relevant to Europa’s radiolytically desiccated surface \citep{trumbo+2019, jabaud+2022, hayes+li2025}. The epsomite and halite samples were acquired as bulk grains from \emph{Sigma-Aldrich}, whereas the water ice samples were created using distilled water. The water was first poured into a silicon container and then vacuum degassed outside of the purge box for approximately $5$ minutes to boil off any atmospheric air trapped within the water. As was standard procedure for all items transferred into the purge box, the distilled water was first placed in the purged glove box antechamber for $30$ minutes to purge residual air and water vapor before moving the sample into the main chamber (see Figure \ref{fig:glovebox_diagram_v1}). After transferring the water container into the main chamber of the glove box, it was held with tongs in a liquid nitrogen dewar to freeze. Once fully frozen (after approximately $5-10$ minutes), the water ice was removed from the container and broken apart into smaller pieces – on the order of approximately $0.5$ cm. The water ice was in the crystalline form.

The following steps were common to all samples of water ice, halite, and epsomite. Each sample was separately ground using an autonomous electric mill (agate mortar and pestle) inside the nitrogen-purged glove box. The mortar was continuously filled with liquid nitrogen to prevent melting of the water ice or alteration of the epsomite and halite, and to ensure fine-grained particulates for all three materials were well-separated during the grinding process, during which liquid nitrogen was continuously poured over the three materials. The grains were wet sieved using liquid nitrogen to further ensure that fine-grained particulates were `rinsed' off of the larger grains, similar to the technique used by \citet{stephan+2021}. The four grain size aliquots are $<63$ $\mu$m (`fine'), 63-250 $\mu$m (`small'), 250-500 $\mu$m (`medium'), $>500$ $\mu$m (`large'). Once a sufficient amount of material was ground and sieved, the grains were collected in $1$ ml nitrogen-cooled vials, closed, and submerged in liquid nitrogen to stay cold. Throughout this process, and especially when working with water ice, liquid nitrogen was continuously poured over all processing equipment to prevent warming and loss of sample.

\begin{deluxetable*}{|l|l|l|l|}
\digitalasset
\tablewidth{0pt}
\tablecaption{Samples measured during this study, along with their recorded temperature range during measurements. Recorded temperatures generally ranged between 100 – 120 K. \label{tab:samples}}
\tablehead{
\colhead{\makecell{Samples Measured - \\Grain Size$\ast$}} & \colhead{\makecell{Temperature Range \\Recorded (K)}} & \colhead{\makecell{Samples Measured - \\Grain Size$\ast$}} & \colhead{\makecell{Temperature Range \\Recorded (K)}}
}
\startdata
\multicolumn{2}{|c|}{Pure Water Ice} & \multicolumn{2}{|c|}{Pure Epsomite (Amb \& Cryo)} \\
\hline
100\% Water Ice - Fine & 112-116 & 100\% Epsomite - Fine & \makecell[l]{Amb: 312-318 \\ Cryo: 111-113} \\
100\% Water Ice - Small & 116-121 & 100\% Epsomite - Small & \makecell[l]{Amb: 314-318 \\ Cryo: 115-117} \\
100\% Water Ice - Medium & 113-117 & 100\% Epsomite - Medium & \makecell[l]{Amb: 311-316 \\ Cryo: 109-111} \\
100\% Water Ice - Large & 110-120 & 100\% Epsomite - Large & \makecell[l]{Amb: 311-316 \\ Cryo: 112-115} \\\cline{3-4}
\multicolumn{2}{|l|}{ } & \multicolumn{2}{|c|}{Pure Halite (Amb \& Cryo)} \\\cline{3-4}
\makecell[l]{50\% Water Ice - Fine \\ 50\% Water Ice - Medium} & 112-113 & 100\% Halite - Fine & \makecell[l]{Amb: 308-319 \\ Cryo: 111-113} \\
\makecell[l]{50\% Water Ice - Small \\ 50\% Water Ice - Large} & 114-116 & 100\% Halite - Small & \makecell[l]{Amb: 297-316 \\ Cryo: 107-109} \\
\makecell[l]{50\% Water Ice - Fine \\ 50\% Water Ice - Large} & 108-109 & 100\% Halite - Medium & \makecell[l]{Amb: 310-315 \\ Cryo: 111-112} \\
\makecell[l]{50\% Water Ice - Small \\ 50\% Water Ice - Medium} & 106-107 & 100\% Halite - Large & \makecell[l]{Amb: 295-305 \\ Cryo: 109-113} \\
\hline
\multicolumn{4}{|c|}{Cross-Material Mixtures} \\
\hline
\multicolumn{2}{|c|}{Epsomite $+$ Water Ice Mixtures} & \multicolumn{2}{|c|}{Halite $+$ Water Ice Mixtures} \\
\hline
\makecell[l]{50\% Epsomite - Large \\ 50\% Water Ice - Large} & 110-112 & \makecell[l]{50\% Halite - Large \\ 50\% Water Ice - Large} & 118-120 \\
\makecell[l]{50\% Epsomite - Large \\ 50\% Water Ice - Fine} & 106-107 & \makecell[l]{50\% Halite - Large \\ 50\% Water Ice - Fine} & 110-112 \\
\makecell[l]{50\% Epsomite - Medium \\ 50\% Water Ice - Medium} & 107-112 & \makecell[l]{50\% Halite - Medium \\ 50\% Water Ice - Medium} & 106-109 \\
\makecell[l]{50\% Epsomite - Medium \\ 50\% Water Ice - Fine} & 107-108 & \makecell[l]{50\% Halite - Medium \\ 50\% Water Ice - Fine} & 112-113 \\
\makecell[l]{50\% Epsomite - Small \\ 50\% Water Ice - Large} & 109-112 & \makecell[l]{50\% Halite - Small \\ 50\% Water Ice - Large} & 115-116 \\
\makecell[l]{50\% Epsomite - Small \\ 50\% Water Ice - Small} & 103-105 & \makecell[l]{50\% Halite - Small \\ 50\% Water Ice - Small} & 105-110 \\
\makecell[l]{50\% Epsomite - Fine \\ 50\% Water Ice - Small} & 107-109 & \makecell[l]{50\% Halite - Fine \\ 50\% Water Ice - Small} & 107-110 \\
\makecell[l]{50\% Epsomite - Fine \\ 50\% Water Ice - Fine} & 107 & \makecell[l]{50\% Halite - Fine \\ 50\% Water Ice - Fine} & 106-107 \\
\makecell[l]{50\% Epsomite - Small \\ 50\% Water Ice - Fine} & 107-108 & \makecell[l]{50\% Halite - Small \\ 50\% Water Ice - Fine} & 112 \\
\enddata
\tablecomments{$\ast$ `Fine' = $<63$ $\mu$m. `Small' = $63-250$ $\mu$m. `Medium' = $250-500$ $\mu$m. `Large' = $>500$ $\mu$m}
\end{deluxetable*}

Intimate mixtures of different grain sizes of water ice, epsomite, and halite were created by hand-mixing, i.e., vigorously shaking, in a $1$ ml vial, then sprinkling onto an LN$_2$-cooled copper sample holder. The material mixtures are all $50\%$/$50\%$ by volume, and were created by first sprinkling $0.5$ ml into the vial, identified visually, and then filling the rest of the vial up with the other material, again visually. Sufficient material was deposited onto the sample holder to ensure that the samples were optically thick. The LN$_2$ dewar that held the sample holder was attached to a linear track that also held a mount for the gold standard – this allowed a smooth and quick transition between measurements of the sample and the gold standard. The reflectance spectra were calculated by ratioing the sample and reference (gold) target measurements. The illumination conditions were identical for the sample and background measurements. The thermocouple was attached to a screw on the sample holder, representing the approximate temperature of the sample itself. The light source was turned on immediately prior to spectral acquisition to reduce thermal changes to the sample. However, the exact temperature of the optical surface of the sample is unknown. Embedding the thermocouple into the sample for each measurement was not feasible for this particular setup, though we estimate that the top of the sample likely increases in temperature perhaps by $10$ -- $20$ K (or even more) as a result of the illumination source; follow-up studies will include a more accurate method for acquiring the true sample temperature. Spectra were collected using a Spectra Vista Corporation HR-1024i NIR spectrometer, with ten spectra taken of each sample and averaged for final spectra. The collection spot size of the spectrometer's fiber-optic cable onto the sample was approximately three centimeters (significantly larger than the largest grain size of material used in this study), and sufficient material was added to the sample holder to ensure that only the sample material was being measured.

\begin{figure*}[ht!]
\plotone{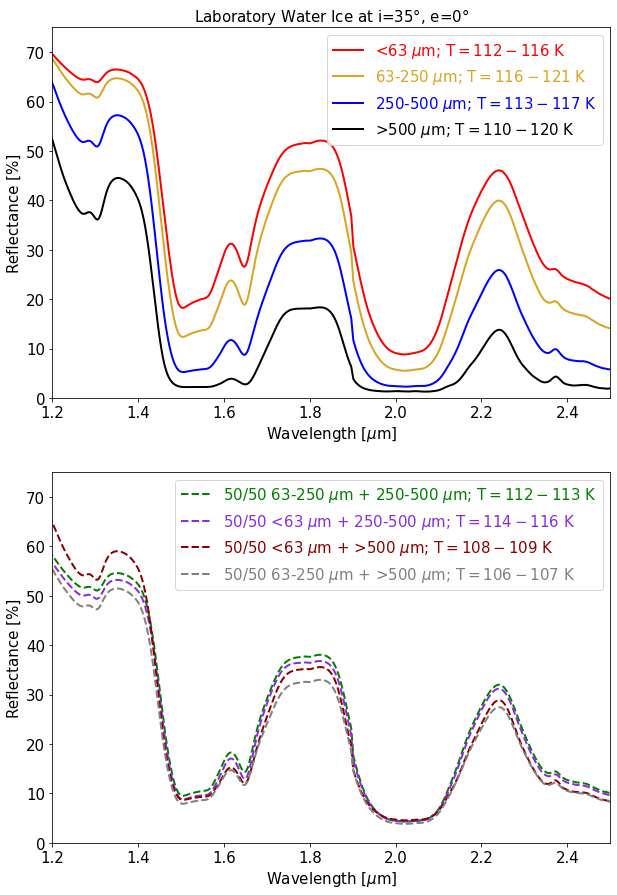}
\caption{Spectra of water ice of the four grain size bins used in this study. \emph{Top}: solid lines represent pure grain size bins. \emph{Bottom}: dashed lines represent $50\%$/$50\%$ by volume mixtures of several combinations of water ice grain sizes. Notice that there is minimal discrepancy between the grain size mixtures, and may therefore be challenging to discriminate visually without spectral mixture analysis. Measured temperatures for each sample are provided in the figure legend as well as in Table \ref{tab:samples}.
\label{fig:labwaterice_puremixes}}
\end{figure*}

\subsection{Spectral Modeling with Hapke Theory}

We used the real and imaginary parts of the index of refraction (\emph{n} and \emph{k}, respectively) for crystalline water ice at $120$ K published by \citet{mastrapa+2008}. Although amorphous water ice optical constants also already exist in the literature, due to the nature of our water ice creation and its temperature during measurements, as well as the prominence of the $1.65$ $\mu$m absorption feature in our spectra that is indicative of the crystalline form of water ice, the crystalline (rather than amorphous) water ice optical constants are the most relevant to this study. We also used the \emph{n} and \emph{k} values published by \citet{dalton+pitman2012} for epsomite at $120$ K. Optical constants for halite in the relevant wavelength range have not been published, likely because in part, halite is relatively bland over this spectral range with only adsorbed water inducing spectral features, and therefore, only water ice and epsomite mixtures were modeled in this study.

Developing synthetic spectra using Hapke theory requires obtaining or estimating values for the porosity, grain size, and temperature, as well as constraining the viewing geometry. The materials produced for the laboratory measurements were confirmed to fall within well-sorted grain size bins of $<63$ $\mu$m, $63-250$ $\mu$m, $250-500$ $\mu$m, and $>500$ $\mu$m due to our careful sample preparation, as laid out in the previous subsection. As described in \citet{hapke2012}, for a distribution of irregularly shaped particles, such as those produced in each of these bins, the representative average grain size, $<D>$, to aid in comparison against models is expected to be of the same magnitude as, but smaller than, the size of the particles, and often corresponds to the smallest grain sizes, rather than to a mean size of the aliquot. \citet{shkuratov+grynko2005} demonstrate that the approximate distribution around the average is $1.0$ for perfect spheres, $0.85$ for rough spheres, $0.6$ for cubes, and $0.2$ for 3D random Gaussian field particles. We chose to approximate our $<D>$ grain size values as $\frac{2}{3}$ of the average grain size of a particular aliquot bin, as our water ice, epsomite, and halite particles likely have shapes somewhere between rough spheres and random Gaussian field particles, and are perhaps slightly closer to that of the rough spheres. Thus, the $<D>$ values used in this study were $21$, $125$, $333$, and $600$ $\mu$m, respectively for the four grain size aliquots.

Each synthetic spectrum uses approximately the same temperature for which the lab-derived spectrum it intends to replicate was acquired. The water ice optical constant data were published with a temperature granularity of $20$ K, i.e., as steps of $20$ K. Optical constants at $120$ K were used for the synthetic spectra for both water ice and epsomite to replicate the likely temperature of the samples themselves (see Table \ref{tab:samples}), as well as the representative temperature of Europa's daytime surface. For the Hapke radiative transfer model, we followed previous parameter selections used by \citet{berdis+2022} and assumed a grain porosity of $0.1$ \citep{johnson+2017} to compute the compaction parameter \citep{hapke1963, buratti1985}, an asymmetry factor of $-0.15$ \citep{buratti1983, buratti1985}, and a mean macroscopic roughness parameter of $10\degree$ \citep{domingue+1991}. We completed a sensitivity analysis to ensure that uncertainties in these parameters would not significantly alter the results; for example, computing modeled spectra of water ice for a grain porosity of $0.5$ instead of $0.1$ did not produce any visibly noticeable differences to the resulting spectra, and therefore we determine that the spectra are not very sensitive to these parameters, even if alternate values would be more applicable. While these parameters were originally chosen for water ice on Europa in \citet{berdis+2022}, we extended their application to epsomite, as there exists far less literature on Hapke application for epsomite. The emission angle and incidence angle used for the modeling matches that used in the laboratory setup, i.e., a $0\degree$ emission angle and a $35\degree$ incidence angle.

The linearly-mixed model was calculated by using Hapke radiative transfer theory to derive the bidirectional reflectance distribution function (BRDF), then `mixing' the two reflectances together. The intimately-mixed model was calculated by again using Hapke radiative transfer theory to calculate the single-scattering albedo (SSA; or \emph{w}), then `mixing' the two SSAs together, and finally, deriving a BRDF reflectance spectrum. It is important to note that both of these models require the use of optical constants, and the traditional `checkerboard' linear mixing approach is not considered here, as it often inherently utilizes laboratory or scenery spectra as its reference.

\begin{figure*}[ht!]
\plotone{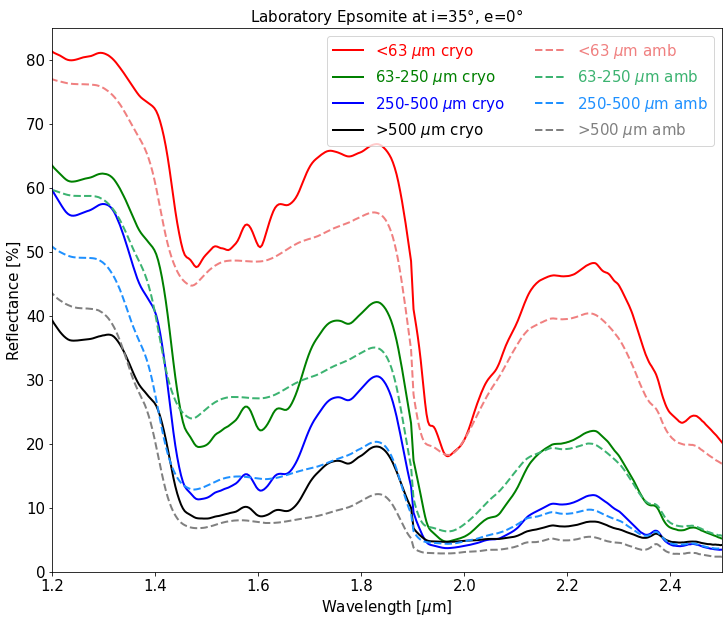}
\caption{Spectra of epsomite of the four grain size bins used in this study, acquired both under ambient (dashed lines) as well as cryogenic (solid lines) conditions. Fine structure in the spectral absorption features in the cryogenic spectra are expected, and are characteristic of colder temperatures. Measured temperatures for each sample are provided in Table \ref{tab:samples}.
\label{fig:labepsomite_ambcryo_pure}}
\end{figure*}

\begin{figure*}[ht!]
\plotone{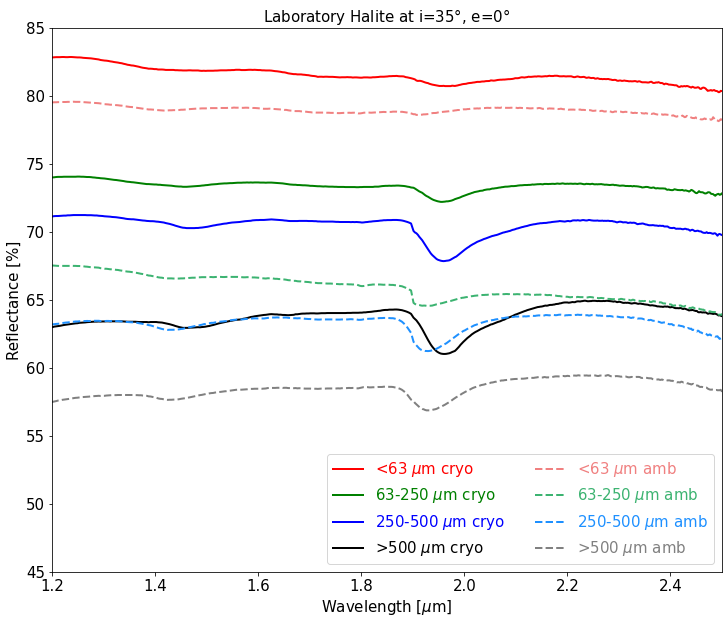}
\caption{Spectra of halite of the four grain size bins used in this study, acquired both under ambient (dashed lines) as well as cryogenic (solid lines) conditions. Fine structure in the spectral absorption features in the cryogenic spectra are expected, and are characteristic of colder temperatures. Measured temperatures for each sample are provided in Table \ref{tab:samples}.
\label{fig:labhalite_ambcryo_pure}}
\end{figure*}

\section{Results}

\subsection{NIR Spectral Measurements of Water Ice, Epsomite, Halite, and Mixtures}

Details of the measurements, including the mixtures that were created and the recorded temperatures during measurements, are presented in Table \ref{tab:samples}. 

Laboratory spectra of pure water ice as single grain sizes and as 50\%/50\% volume mixtures are provided in Figure \ref{fig:labwaterice_puremixes}, and recorded temperatures range from $106 - 120$ K. Characteristic absorption features of water ice are present at $1.25$, $1.5$, $1.65$, and $2.0$ $\mu$m. An artifact exists in many of our lab measurements, and is noteworthy in the water ice and epsomite spectra; a small reflection peak occurring at $\sim2.38$ $\mu$m is likely a result of an organic contamination on the gold standard used in this study, and should be largely ignored. The feature exists as an absorption feature in the spectrum of the gold standard, and does not exist in the spectrum of the samples - we therefore attribute this to a carbon-based blemish on the surface of the gold standard that, when the reference and target spectra are ratioed together, appears as a peaked feature in the resulting reflectance spectrum. As grain sizes increase, the reflectance at all wavelengths decreases, and the absorption features reach a minimum (i.e., saturate), consistent with previous works, e.g., \citet{clark1981} and \citet{stephan+2021}. Based on the absorption depth of the $1.65$ $\mu$m feature, which is highly dependent on temperature \citep{grundy+schmitt1998}, the temperatures recorded above in Table \ref{tab:samples} are consistent with the temperatures of the samples themselves, likely to within approximately $10-20$ K. Note that the presence of the $1.65$ $\mu$m absorption band, and the shape and position of the $2.0$ $\mu$m absorption band, demonstrate that the water ice is indeed in the crystalline form and is not too different from the thermocouple-measured temperature.

\begin{figure*}[ht!]
\plotone{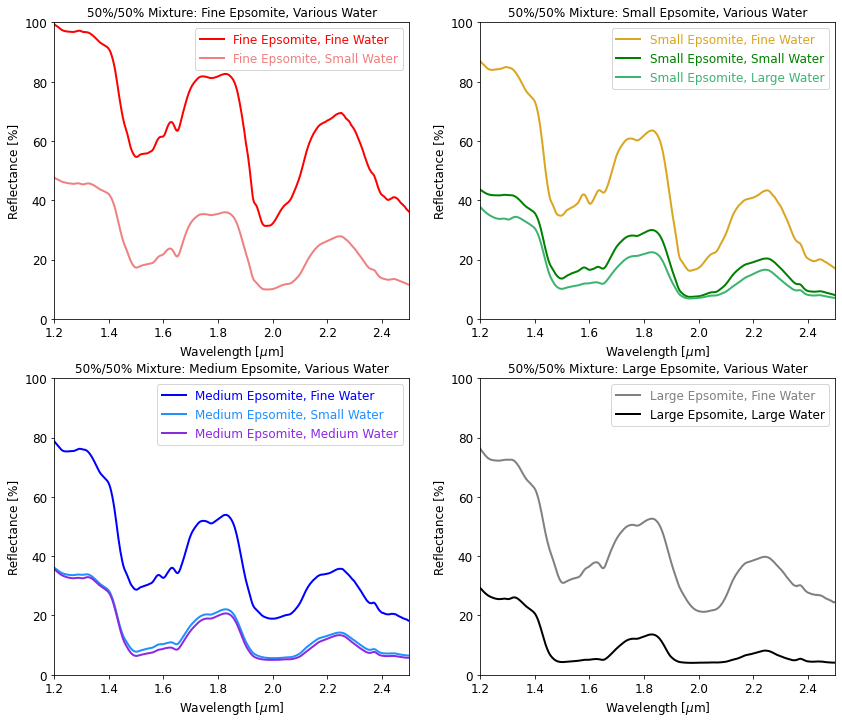}
\caption{50\%/50\% vol. mixtures of epsomite and water ice for different grain aliquots, where `fine' is $<63$ $\mu$m, `small' is 63-250 $\mu$m, `medium' is 250-500 $\mu$m, and `large' is $>500$ $\mu$m. The reader is referred to Figure \ref{fig:labwaterice_puremixes} of pure water ice as reference to see the influence that water ice has on the mixed spectra shown in this figure.
\label{fig:lab_epsomite_water_mixtures_v2}}
\end{figure*}

\begin{figure*}[ht!]
\plotone{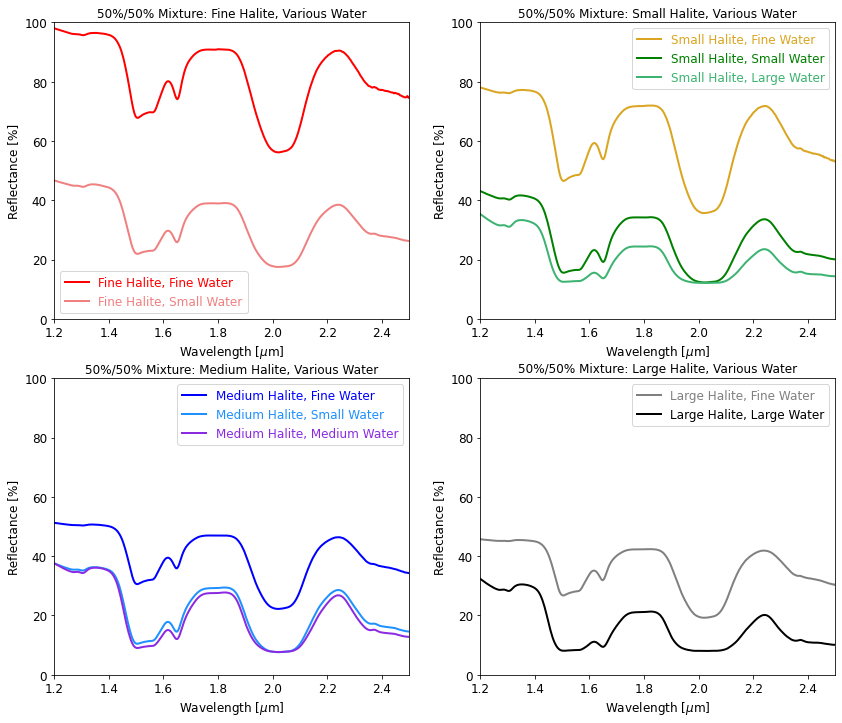}
\caption{50\%/50\% vol. mixtures of halite and water ice for different grain aliquots, where `fine' is $<63$ $\mu$m, `small' is 63-250 $\mu$m, `medium' is 250-500 $\mu$m, and `large' is $>500$ $\mu$m. The reader is referred to Figure \ref{fig:labwaterice_puremixes} of pure water ice as reference to see the influence that water ice has on the mixed spectra shown in this figure.
\label{fig:lab_halite_water_mixtures_v2}}
\end{figure*}

Laboratory spectra of epsomite (Figure \ref{fig:labepsomite_ambcryo_pure}) and halite (Figure \ref{fig:labhalite_ambcryo_pure}) were acquired both at cryogenic conditions as well as ambient conditions. The ambient spectra were acquired first, then the liquid nitrogen was added to the dewar to lower the temperature of the sample for the cryogenic measurements. Cryogenic spectra of epsomite exhibited a plethora of absorption features not present in the ambient spectra. Cryogenic and ambient halite both showed a feature near $1.95$ $\mu$m. This is in agreement with previous literature of published epsomite and halite spectroscopic data (e.g., \citealt{deangelis+2017} and \citealt{hanley+2014}, respectively).

The $50\%$/$50\%$ mixtures of epsomite and water ice (Figure \ref{fig:lab_epsomite_water_mixtures_v2}) and halite and water ice (Figure \ref{fig:lab_halite_water_mixtures_v2}) demonstrate the dominating impact that water ice has on the spectra. As grain sizes of each material increases, we see reduced overall reflectances and saturated absorption features. As an example, for extreme mixtures, such as `Large Epsomite, Fine Water', the influence of the higher reflectance and thus larger contrast in the absorption features in the fine grains of water ice dominates the mixed spectrum, with only minimal epsomite absorption features being observable. This effect has been noticed previously, specifically, ``when there is a large disparity in the sizes of the components of an intimate mixture, the fine particles can have an effect all out of proportion to their mass fraction'' \citep{hapke1993}. Mixtures of small and medium grain sizes of epsomite and water ice are particularly challenging to discern from each other. The mixtures of halite and water ice are almost indiscernible from pure water ice. The prominent NIR spectral feature of halite at $1.95$ $\mu$m is barely detectable in mixtures with water ice, and the influence of fine-grained halite exhibits a slight influence on the left side of the $2.0$ $\mu$m water ice feature, causing an asymmetry in the absorption feature.

\subsection{Modeling of the NIR Spectral Measurements}

\begin{figure*}[ht!]
\plotone{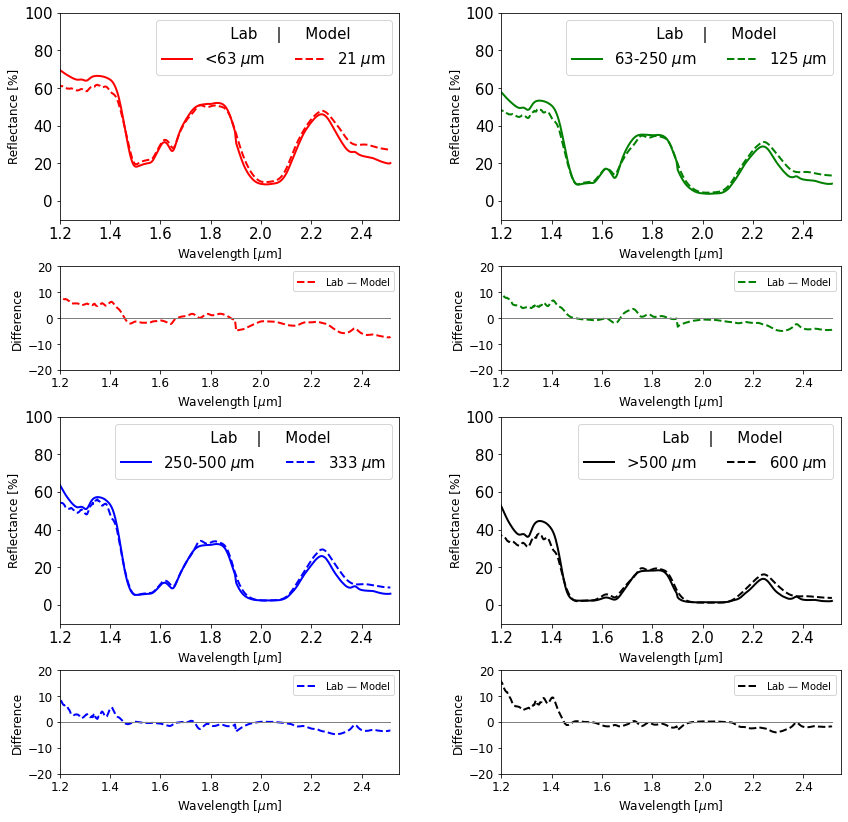}
\caption{Model (dashed) and laboratory (solid) spectra for individual grain sizes of water ice are generally well-matched. Dashed line appears solid shortward of $1.4$ $\mu$m due to noise. Model grain sizes are $\frac{2}{3}$ of the aliquot average, $<D>$. Differences between laboratory and modeled spectra are provided in the ``Difference'' plots below each corresponding spectrum, and generally remain within $\sim10\%$.
\label{fig:pureH2O_labvsmodel_diff}}
\end{figure*}

We first demonstrate the validity and limitations of our experimental effort by comparing modeled water ice spectra to the reflectance of the water ice spectra we obtained. Our laboratory spectra of pure water ice compared to their Hapke modeled spectra using the average grain size, $<D>$, as described above, is shown in Figure \ref{fig:pureH2O_labvsmodel_diff}. The largest variations between the laboratory and modeled water ice spectra appear to exist in the continua shortward of $1.45$ $\mu$m and longward of $2.3$ $\mu$m. This may indicate that future spectral mixture analyses may require spectrally localized modeling of the $1.45-2.3$ $\mu$m region where the $1.5$ $\mu$m and $2.0$ $\mu$m water ice absorption features exist if the intent of the modeling is to perform a high fidelity compositional analysis of water ice and other related hydrated species.

\begin{figure*}[ht!]
\plotone{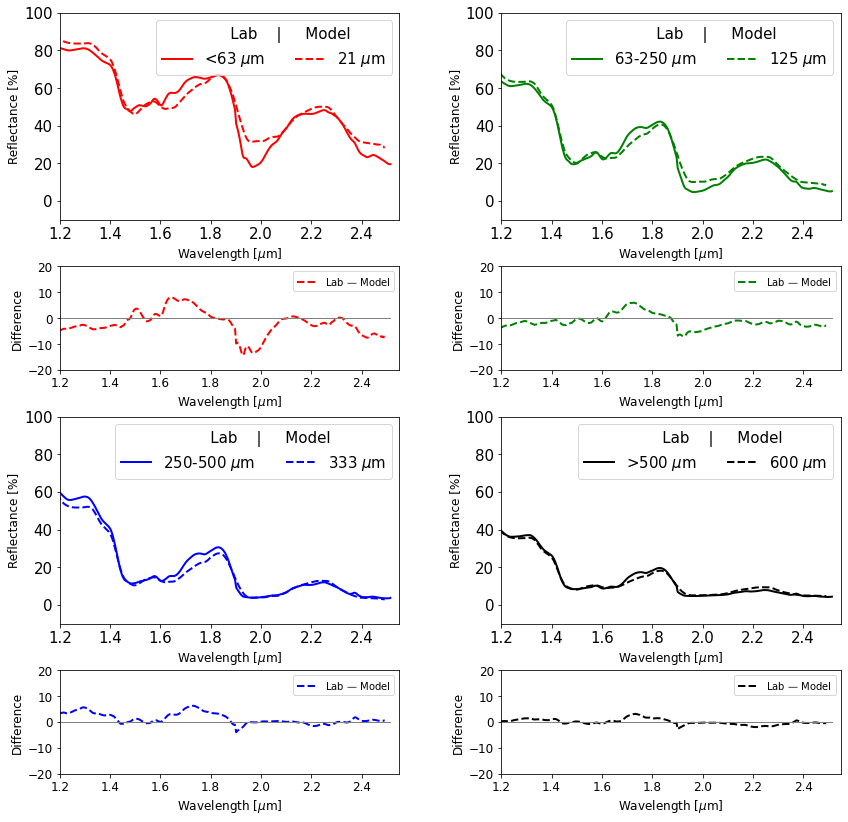}
\caption{Model (dashed) and laboratory (solid) spectra for individual grain sizes of epsomite are generally well-matched, though not as well as that of water ice. Dashed line appears solid shortward of $1.4$ $\mu$m due to noise. Model grain sizes are $\frac{2}{3}$ of the aliquot average, $<D>$. Differences between laboratory and modeled spectra are provided in the ``Difference'' plots below each corresponding spectrum, and generally remain within $\sim10\%$.
\label{fig:pureEPS_labvsmodel_diff}}
\end{figure*}

Similarly, laboratory spectra of separate grain size aliquots of epsomite compared to their Hapke modeled spectra (Figure \ref{fig:pureEPS_labvsmodel_diff}) show general similarities, but also significant spectral differences, certainly greater than for water ice. Differences between the laboratory and modeled epsomite are particularly striking in the slope between $1.6-1.8$ $\mu$m, in the spectral features of that region, and in the shape of the absorption feature at $2.0$ $\mu$m. Furthermore, the modeled fits to the laboratory data appear more similar to the laboratory data for the larger grain sizes than for the smaller grain sizes. The modeled data of epsomite appear more similar to room temperature epsomite than they do to cryogenic epsomite.

\begin{figure*}[ht!]
\plotone{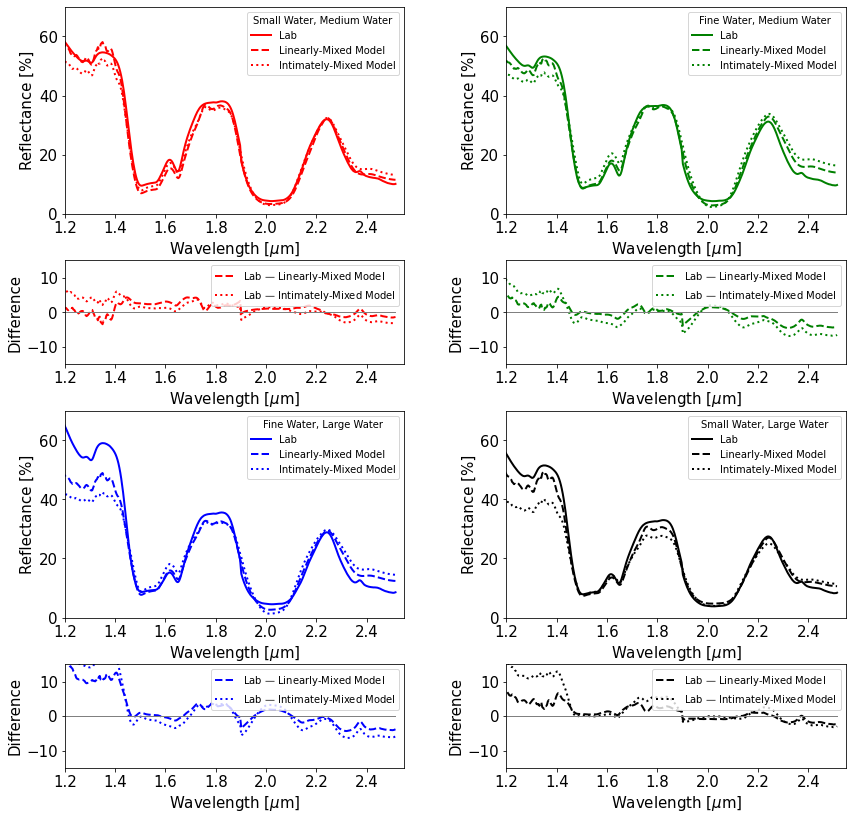}
\caption{Linearly-mixed model (dashed), intimately-mixed model (dotted), and laboratory (solid) spectra for $50\%$/$50\%$ volume mixtures of different grain sizes of water ice. Differences between laboratory and modeled spectra are provided in the ``Difference'' plots below each corresponding spectrum, and generally remain within $\sim10\%$.
\label{fig:lab_linmix_intmix_diff_waterice}}
\end{figure*}

Laboratory and modeled mixtures of water ice grains are provided in Figure \ref{fig:lab_linmix_intmix_diff_waterice}. As with the singular aliquots of water ice, the differences between the model and laboratory spectra for the smaller grain sizes (i.e., the top half of Figure \ref{fig:lab_linmix_intmix_diff_waterice}), are within $\sim10\%$ for both the linearly- and intimately-mixed models. However, the model differentials, calculated by subtracting the modeled spectrum from the laboratory spectrum, rise above $10\%$ for the larger grain sizes, and are more apparent in the intimately-mixed model.

\begin{figure*}[ht!]
\plotone{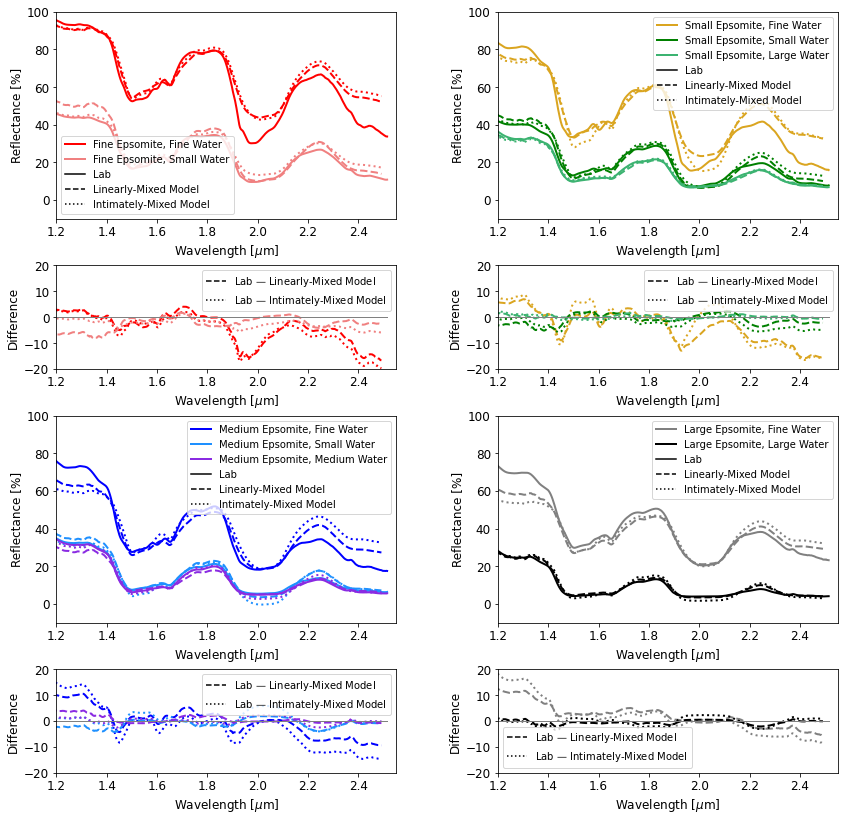}
\caption{Linearly-mixed model (dashed), intimately-mixed model (dotted), and laboratory (solid) spectra for $50\%$/$50\%$ volume mixtures of different grain sizes of water ice and epsomite. Differences between laboratory and modeled spectra are provided in the ``Difference'' plots below each corresponding spectrum. Similar to the pure epsomite spectra, the modeled data does not exhibit as fine structure in the absorption features that are expected, and observed, in the laboratory data. Differences generally remain within $\sim15\%$.
\label{fig:lab_linmix_intmix_diff_water-eps}}
\end{figure*}

Laboratory and modeled mixtures of water ice and epsomite grains are provided in Figure \ref{fig:lab_linmix_intmix_diff_water-eps}. Both linear and intimate mixtures again perform similarly. Also, similar to the singular aliquots of epsomite, there is a significant difference, again calculated by subtracting the modeled spectrum from the laboratory spectrum, between the model and laboratory spectra in the $1.5$ $\mu$m and $2.0$ $\mu$m bands, and in the continuum at $2.2-2.45$ $\mu$m. Modeled spectra of epsomite mixed with water ice contain very little influence of the epsomite, despite being $50\%$/$50\%$ mixtures. These modeled spectra are consistently $\sim10\%$ darker at all wavelengths compared to the laboratory data, and have been shifted up in these figures to allow for better visual comparison. Particularly evident in the smaller grain sizes, i.e., the top half of Figure \ref{fig:lab_linmix_intmix_diff_water-eps}, is a triple-lobed feature in the lab data at the $2.0$ $\mu$m band, compared to a more smoothed, rounded absorption feature in the modeled data. \citet{mccord+2001} likened the source of this fine structure to the fact that, in epsomite and other magnesium sulfate hydrates, ``fine structure appears as cold temperatures reduce thermal broadening in water absorption features.'' However, these features are not in the optical constants of cryogenic epsomite from \citet{dalton+pitman2012}. Both the linearly- and intimately-mixed model results of water ice and epsomite mixtures do not appear to be greatly affected by the epsomite. For example, the shapes of the $1.5$ $\mu$m and $2.0$ $\mu$m absorption features appear to more closely match the shape of water ice absorption features at those wavelengths, or even the ambient spectra of epsomite. This effect is more prominent for smaller grain sizes of water ice, and in fact, smaller grain sizes of each material tend to have a stronger influence on the spectrum; this occurs with the epsomite as well, as its primary absorption features are largely influence by the H$_2$O attached to the magnesium sulfate in epsomite.
\section{Discussion}

\subsection{Water Ice Lab Spectra and Model Fits}

The water ice grain size mixtures are relatively well-represented (within $5\%$ reflectance at a majority of wavelengths) by the linearly- and intimately-mixed model. Minor differences in the spectra may result from a modeled $<D>$ grain size that is not adequately representative of the laboratory grain aliquots, however, we find that while altering the modeled $<D>$ grain size does indeed influence the resulting spectra, it does not account for the inability to match both the continuum and the absorption features across the whole wavelength region that is modeled. As mentioned previously, the linearly- and intimately-mixed models can adequately recreate the absorption features \emph{or} the continua, but not both. The intimately-mixed modeling approach is expected to produce improved fits, however, it does not appear to be as significant as anticipated. The intimately-mixed model displays a less steep slope in the continuum across the full wavelength range, however this may be a result of the acquisition techniques of the laboratory spectra, rather than an inability of the intimately-mixed model to fit the data. In the regions of interest for water ice, i.e., the $1.5$ $\mu$m and $2.0$ $\mu$m regions, the linearly-mixed model and the intimately-mixed model perform comparably at fitting the laboratory data. While this result is unexpected, it has been observed in previously published studies; \citet{stack+milliken2015} found that intimate and linear model mixtures perform adequately for clay and sulfate mixtures, and in fact, claim that the linear model performed slightly better than the intimate model at reproducing the laboratory measurements.

\subsection{Epsomite Lab Spectra and Model Fits}

The spectral features in epsomite are likely due to the narrowing and separation of multiple water-related absorption features at cryogenic temperatures that otherwise merge into one broad feature at ambient temperature (e.g., \citealt{mccord+2001, dalton+pitman2012}). Figure \ref{fig:pureEPS_labvsmodel_diff} demonstrates that even without mixing in the water ice, pure epsomite optical constants do not adequately model the laboratory data. There are several hypotheses for why this might be the case: (1) the epsomite optical constants in \citet{dalton+pitman2012} may have been incorrectly labeled as `cryogenic' when archived; and/or (2) a portion of our laboratory epsomite desiccated into hexahydrite (MgSO$_4$ $\cdot$ $6$H$_2$O), which is likely, as \citet{omori+kerr1963} find that epsomite on exposure at ordinary conditions dehydrates easily to hexahydrite. However, the cryogenic epsomite optical constants published in \citet{dalton+pitman2012} are identical to the ambient optical constants, with the exception of a small reflectance shift, and this is not consistent with results published in their Figure 1.

\subsection{Halite Lab Spectra}

Adsorbed water onto halite is likely inducing the absorption feature at $1.95$ $\mu$m, and is not intrinsic to halite, which is known to have no spectral features in this wavelength range. The mixtures of halite and water ice are almost indiscernible from pure water ice as expected, and, halite being transparent, is highly scattering, ensuring water ice particles strongly interact with the incident illumination. For mixtures of fine-grained halite and water ice, the greater band depth at $1.95$ $\mu$m may be due to the fine halite grains coating the water ice grains and thus dominating the spectrum. Or, this larger absorption feature is simply due to the fact that fine-grained particles have a larger surface area, and therefore more surface-adsorbed water is present (e.g., \citealt{dyar+2010}).

\subsection{Implications and Future Work}

The MISE instrument aboard Europa Clipper, as well as the MAJIS instrument aboard JUICE, will acquire and return hyperspectral imaging data from $0.8-5.0$ $\mu$m and $0.5-5.56$ $\mu$m respectively, with spatial resolutions as low as tens of meters per pixel \citep{blaney+2024, poulet+2024}. Significant surface compositional breakthroughs resulted from spectral mixture analyses performed on Galileo's hyperspectral imager, the Near-Infrared Mapping Spectrometer (NIMS; \citealt{carlson+1992}), however, failure to fully deploy the high gain antenna on Galileo unfortunately resulted in limited data transmission, which particularly impacted data return in the $\sim 3 - 5$ $\mu$m range \citep{johnson1994}. It is anticipated that not only will the MISE and MAJIS instruments be able to transmit data in this critical wavelength range, but will also provide significantly improved resolution, both spectrally and spatially. Spectral mixture analyses across the surface of Europa using data from these instruments can therefore lead to even more exciting compositional breakthroughs. 

Europa’s surface likely contains areas of intimate and linear mixtures of materials (e.g., \citealt{spencer1987, hansen+mccord2004, hibbitts2023}), and both models may be applicable to future spectral mixture analyses that use Europa Clipper MISE and JUICE MAJIS data. Particularly noteworthy in this study is the fact that a spectrum of $50\%$/$50\%$ mixtures of two grain sizes (of either water ice, epsomite, halite, or a mixture) does not exhibit a reflectance spectrum ``half-way'' between its two constituent pure spectra, therefore demonstrating the need for  quantitative rather than qualitative mixture analyses for abundance determinations, i.e., in the form of a spectral mixture analysis that uses optical constant data. Furthermore, a mixture of fine and medium grain sizes should have a higher overall reflectance than a mixture of small and medium, but this is not observed in the laboratory data (e.g., Figure \ref{fig:labwaterice_puremixes}). If this result is correct, it indicates that significant observational bias may exist when identifying species and/or grain sizes by their spectral characterization, and mitigation could include performing a high fidelity spectral mixture analysis with optical constants to ensure accurate abundances of materials and their grain sizes are determined.

A follow-on study to the work presented here will include deriving updated epsomite optical constants, and several of these hypotheses concerning the discrepancies between the laboratory spectra of epsomite and the modeled spectra of epsomite will be tested. However, current spectral mixture analyses that use these epsomite optical constant data should be aware of this likely error in the published data, and derived abundances may not be fully representative.

\section{Conclusions}

We publish the first NIR spectra of grain particulate mixtures of water ice, epsomite, and halite at cryogenic temperatures. Furthermore, we performed a quantitative assessment of the ability of both intimately- and linearly-mixed models to reproduce laboratory data of different grain mixtures of water ice, as well as water ice mixed with epsomite. We find that the linearly-mixed and intimately-mixed models of water ice appear to match the laboratory spectra as expected, though still display some inconsistencies, often either in the continuum or the absorption features. When modeling pure water ice, no discernable difference is observed between the linearly- and intimately-mixed models. Spectral mixture analyses that use water ice optical constants should perform both linear and intimate unmixing processes to identify a range of potential abundance uncertainty. 

We find that smaller grains of water ice impart a stronger influence than the larger grains of water ice on the $2.0$ $\mu$m spectral feature of epsomite, and grain size signatures for both halite and epsomite are challenging to discern for larger grain sizes as a result of the saturated absorption features. This may indicate that an observation bias toward smaller grain sizes could exist, and that quantitative assessments provided by spectral mixture analyses will be the most reliable method for determining compositions and abundances of materials on Europa's surface. 

We also find that model fidelity of mixtures with epsomite are inconclusive due to the potentially incorrectly published epsomite optical constant data. Future spectral mixture analyses that include epsomite should be aware of a potential error in the published epsomite optical constant data, in which the data labeled as `cryogenic' appears to be taken at ambient conditions. In a follow-on study, we intend to publish corrected optical constant data for epsomite and pursue the fidelity of intimately- and linearly-mixed modeling of laboratory epsomite data.

\section{Data Availability}

The spectral data presented in this study can be made fully available upon request to the corresponding author.


\begin{acknowledgments}
This work was supported by NASA through the MISE investigation of the Europa Clipper project. A special thanks to the two anonymous reviewers whose feedback improved the quality of this manuscript.
\end{acknowledgments}

\bibliography{main}{}
\bibliographystyle{aasjournalv7}



\end{document}